\documentstyle[aps,pre,psfig,preprint,tighten]{Revtex}
\begin{document}
\draft

\title{Microscopic expressions for the thermodynamic temperature}
\author{Owen G. Jepps, Gary Ayton and Denis J. Evans}
\address{Research School of Chemistry \\ The Australian National
University \\ Canberra, ACT 0200 \\ Australia}
\date{\today}
\maketitle

\begin{abstract}
We show that arbitrary phase space vector fields can be used to
generate phase functions whose ensemble averages give the
thermodynamic temperature. We describe conditions for the validity of
these functions in periodic boundary systems and the Molecular
Dynamics (MD) ensemble, and test them with a short-ranged potential MD
simulation.
\end{abstract}


\section{Introduction}
\label{sec:intro}

The temperature of an equilibrium system is calculated from the mean
kinetic energy of its particles. However, Rugh \cite{rugh} has
recently derived a new expression whose average yields the temperature
in the microcanonical ensemble. The derivation involves the
differentiation of the phase space volume of the ensemble (whose
logarithm gives the entropy) with respect to the energy. The resultant
expression depends not only on the momenta of the particles, but also
on their (spatial) configuration. This result reflects the fact that
the temperature of a system effects the configurations adopted by that
system, and consequently one would expect to be able to calculate the
temperature from configurational as well as kinetic information.

In \cite{brent}, a completely configurational version of Rugh's result
was derived, and applied to Monte Carlo simulations, where the
equipartition theorem cannot be used since only the configurational
degrees of freedom are considered \cite{allen}. Not only did the
temperature indeed correspond to the input temperature, but it was
found to be a useful diagnostic for revealing coding errors as well.
In \cite{gary}, Rugh's result was applied to the non-equilibrium
domain, as an alternative means of defining the local temperature. As
such, it correctly accounted for heat fluxes where the kinetic
temperature failed to do so.

These applications have required extensions of Rugh's original work.
In this paper we will justify these modifications. We
will prove that the temperature expression used in \cite{gary} is
indeed equal to Rugh's expression to $O(1/N)$, and that the
configurational temperature expression given by Eq.\ (8) in
\cite{brent} holds. 

For practical reasons, atomistic simulations are usually conducted
under periodic boundary conditions. Furthermore, linear momentum is
conserved in molecular dynamics (MD) simulations. Thus the ensembles
explored by MD simulations are not the full microcanonical or
canonical ensembles, but subsets of these (the ``MD ensembles''). We
will therefore also prove that the temperature expressions in
\cite{brent,gary} hold for periodic boundary conditions, and in the
canonical and microcanonical MD ensembles. 

In order to prove these results, we will derive a more general
temperature expression in Sec. \ref{sec:emget}. Within the framework
of this broader result, we will consider which functions yield the
temperature for periodic systems and MD simulations. In Sec.
\ref{sec:formulae} we will use this theorem to obtain more
specific results such as the equipartition theorem and Rugh's result.
Finally, in Sec. \ref{sec:example} we will apply these
temperature-yielding functions to MD simulations of systems with
short-range potentials.

\section{Generalised Temperature Expressions}
\label{sec:emget}

Let us consider an $N$-particle system at equilibrium. We define
$\bbox{\Gamma}=(\Gamma_1,\ldots,\Gamma_{6N})=(p_1,\ldots,p_{3N},q_1,\ldots,q_{3N})$,
where the $q_i$ and $p_i$ represent the $6N$ spatial coordinates and
conjugate momenta which determine the dynamics of the system via
Hamilton's equations. The energy of our system is given by the
Hamiltonian ${\cal H}(\bbox{\Gamma}) = \sum p_i^2/m + V(\{q_j\})$,
where $V$ represents the potential energy of the system.

In this section we will prove the following result. Suppose we choose
a vector field ${\mathbf B(\Gamma)}$ such that
$0 < |\left\langle \nabla{\cal H}\cdot{\mathbf B(\Gamma)} \right\rangle| < \infty$,
$0 < |\left\langle \nabla \cdot {\mathbf B(\Gamma)} \right\rangle| < \infty$ 
 (where $\left\langle \ldots\right\rangle$ represents an ensemble average) and
$\left\langle \nabla{\cal H}\cdot{\mathbf B(\Gamma)} \right\rangle$ grows
more slowly than $e^N$ in the thermodynamic limit. Then
\begin{equation}
\label{EMGET:gen}
\frac{ \left\langle \nabla{\cal H}\cdot{\mathbf B(\Gamma)} \right\rangle }
     { \left\langle \nabla \cdot {\mathbf B(\Gamma)} \right\rangle       }
 = kT.
\end{equation}
This result, as applied to the canonical ensemble, appears in \cite{allen}
without proof. For certain choices of ${\mathbf B(\Gamma)}$, it is possible 
to derive such temperature expressions using
the approach of Gray \& Gubbins \cite{gray&gubbins} --- however, it is not
evident how to extend their method of deriving hypervirial relations 
to arbitrary ${\mathbf B(\Gamma)}$.

In Sec. \ref{sub:micro} and \ref{sub:canon} we will prove
Eq.\ (\ref{EMGET:gen}) for the microcanonical and canonical ensembles 
respectively. Moreover, it is possible to apply Eq.\ (\ref{EMGET:gen}) to 
systems with periodic boundary conditions and the ``MD ensembles''. We will 
develop the additional necessary conditions in Sec. \ref{sub:micro:period} 
and \ref{sub:canon:period}.

\subsection{Microcanonical ensemble}
\label{sub:micro}

We begin with a proof of Eq.\ (\ref{EMGET:gen}) in the microcanonical
ensemble. Consider our $N$-particle system, whose physical size is
determined by a set of barriers or walls. If we denote by $\Omega$ the
set of all allowed $\bbox{\Gamma}$ within our phase space, then the
extent of $\Omega$ in the spatial coordinates is limited by the
physical size of the system. The momenta are unbounded, so that
$\Omega$ forms a cylinder in phase space.

We define the surface of constant energy $A(E) := \{\bbox{\Gamma} :
{\cal H}(\bbox{\Gamma}) = E \}$, and the set of all points of equal or
lower energy $\Omega(E):=\{\bbox{\Gamma}:{\cal H}(\bbox{\Gamma})\le
E\}$. Thus $\Omega = \Omega(\infty)$. Traditionally, the
microcanonical ensemble is the set of phase points whose energy lies
between $E$ and $E+\Delta E, \Delta E \ll E$, with each point being
equally likely to occur. Thus our (dimensionless) partition function
would be
\[
\frac{1}{h^{3N}N!}\int\limits_{\Omega(E+\Delta E) \backslash
\Omega(E)} \;{\sf d}{\bbox{\Gamma}},
\]
where $h$ is Planck's constant. In the limit as $\Delta E \rightarrow
0$, the microcanonical ensemble of energy $E$ becomes the surface
$A(E)$ with a probability distribution given by the partition function
\[
\frac{1}
     {h^{3N}N!} 
\int\limits_{\Omega} 
   \delta({\cal H}(\bbox{\Gamma})-E) \;{\sf d}{\bbox{\Gamma}} 
= 
\frac{1}
     {h^{3N}N!} 
\int\limits_{A(E)} 
   \frac{\;{\sf d}{A_E}}{\|\nabla {\cal H}(\bbox{\Gamma})\|}
= 
\frac{1}{h^{3N}N!} \int\limits_{A(E)} {\sf d}\mu_{E},
\]
where ${\sf d}{A_E}$ represents the infinitesimal area element on
$A(E)$ at $\bbox{\Gamma}$, and ${\sf d}\mu_{E}={\sf d}{A_E}/\|\nabla
{\cal H}(\bbox{\Gamma})\|$. We note immediately the dimensional
inconsistency of this expression. From a physical point of view, it is
necessary at this point to reduce the spatial coordinates and momenta
to dimensionless units $p_i^*$ and $q_i^*$. Consequently, all
functions of these coordinates will remain dimensionless, and can be
converted to the appropriate units after calculation in reduced units.
Physically, this is equivalent to selecting a standard basis of units
by which to measure all the $p_i$, and all the $q_i$ (and
consequently, the units of our resultant temperature). However, this
choice of units for the $p_i$ and $q_i$ can always be made
independently of each other, so that this does not present a problem
in practice. Indeed, since there is no unique choice of units for
converting the phase space coordinates to a reduced form, there will
be infinitely many different instantaneous expressions for the
temperature which will all give equivalent values in different bases of 
units (see, e.g., \cite{morriss}).
In what follows, we will assume that our $p_i$ and $q_i$ are dimensionless as
required (as will be the factor $h^{3N}$).

In the surface ensemble, the entropy $S(E)$ can be defined as
\[
e^{S(E)/k} = \frac{1}{h^{3N}N!} \int\limits_{A(E)} {\sf d}\mu_{E},
\]
and the average value of a phase function ${\cal B}(\bbox{\Gamma})$ in
the ensemble is 
\[
\left\langle {\cal B}(\bbox{\Gamma}) \right\rangle_E =
\frac{\int_{A(E)} {\cal B}(\bbox{\Gamma}) {\sf d}\mu_{E}}{\int_{A(E)}
{\sf d}\mu_{E}} = 
\frac{e^{-S(E)/k}}{h^{3N}N!} \int\limits_{A(E)} {\cal
B}(\bbox{\Gamma}) {\sf d}\mu_{E}.
\]
The temperature of a microcanonical ensemble with energy $E$,
temperature $T$, entropy $S$, and volume $V$ is determined via the
thermodynamic relation
\begin{equation}
\label{def:temp}
\frac{1}{T(E)} = \left. \frac{\partial S(E)}{\partial E} \right|_{V}.
\end{equation}
Suppose, for an arbitrary vector field ${\mathbf B(\Gamma)}$, we define
${\cal B}(\bbox{\Gamma}) = \nabla{\cal H}(\bbox{\Gamma})\cdot{\mathbf B(\Gamma)}$
and
\begin{equation}
\label{def:WB}
W_{{\cal B}}(E) = e^{S_{{\cal B}}(E)/k} 
                 = \frac{1}{h^{3N}N!} \int\limits_{A(E)} {\cal
B}(\bbox{\Gamma})\ {\sf d}\mu_{E} ,
\end{equation}
where we assume $0<\left\langle {\cal
B}(\bbox{\Gamma}) \right\rangle_E<\infty$. From first principles, 
\begin{eqnarray*}
h^{3N}N!\ 
\frac{\partial W_{{\cal B}}(E)}
     {\partial E} 
&=&
\lim_{\delta\rightarrow0} 
h^{3N}N!\ 
\frac{W_{{\cal B}}(E+\delta)-W_{{\cal B}}(E)}
     {\delta} \\
&=&
\lim_{\delta\rightarrow0} 
\frac{1}
     {\delta} 
\left[ 
   \int\limits_{A(E+\delta)} 
         {\cal B}(\bbox{\Gamma})\ {\sf d}\mu_{E+\delta}
-
\int\limits_{A(E)} 
         {\cal B}(\bbox{\Gamma})\ {\sf d}\mu_{E}  \right] \\
&=&
\lim_{\delta\rightarrow0} 
\frac{1}
     {\delta} 
\left[ 
\int\limits_{A(E+\delta)} 
    {\mathbf B(\Gamma)} \cdot \hat{\mathbf{n}}(\bbox{\Gamma}) 
    \;{\sf d}{A}_{E+\delta} 
-
\int\limits_{A(E)} 
    {\mathbf B(\Gamma)} \cdot \hat{\mathbf{n}}(\bbox{\Gamma}) 
    \;{\sf d}{A}_E \right],
\end{eqnarray*}
where $\hat{\mathbf{n}}(\bbox{\Gamma})$ is a unit normal vector to the
surface $A(E)$ at $\bbox{\Gamma}$. We apply Gauss' Theorem to obtain
\begin{eqnarray*}
h^{3N}N!\ \frac{\partial W_{{\cal B}}(E)}{\partial E} 
&=&
 \lim_{\delta\rightarrow0} \frac{1}{\delta} 
\int\limits_E^{E+\delta} \int\limits_{A(\xi)} 
     \nabla \cdot {\mathbf B(\Gamma)} \ \;{\sf d}{A_\xi} \;{\sf
d}{\xi} \\
&=&
 \int\limits_{A(E)} \nabla \cdot {\mathbf B(\Gamma)} \ {\sf d}\mu_{E}.
\end{eqnarray*}
Therefore it follows that
\begin{equation}
\label{EMGET:microcanonical}
\frac{1}{kT_{{\cal B}}(E)} := \frac{1}{k} \frac{\partial S_{{\cal
B}}(E)}{\partial E} = 
\frac{ \left\langle \nabla \cdot {\mathbf B(\Gamma)} \right\rangle_E
}{ \left\langle \nabla{\cal H}\cdot{\mathbf B(\Gamma)} \right\rangle_E
}.
\end{equation}
From Eq.\ (\ref{def:WB}), we have that $S_{{\cal B}}(E) = S(E) + 
k \ln \left\langle {\cal B}(\bbox{\Gamma}) \right\rangle_E$. 
Now, as long as $\left\langle {\cal B}(\bbox{\Gamma}) \right\rangle_E$
grows more slowly than $e^N$ in the thermodynamic limit,
$\partial S_{{\cal B}}(E)/\partial E  = \partial S(E)/\partial E$
 in the thermodynamic limit (a relation we will henceforth denote as 
$\partial S_{{\cal B}}(E)/\partial E \simeq \partial S(E)/\partial E$).
Consequently, $T_{{\cal B}}(E) \simeq T(E)$, and we recover the same
result as Eq.\ (\ref{EMGET:gen}). Furthermore, we may drop the condition
that $\left\langle {\cal B}(\bbox{\Gamma}) \right\rangle_E$ be positive,
since Eq.\ (\ref{EMGET:microcanonical}) gives the same value, whether we
use $\pm \mathbf B(\Gamma)$.

In order to apply Gauss' Theorem, we require that
$\nabla{\cal H}$ be continuous (ie that ${\cal H}$ be differentiable) 
for finite energies. As far as conditions on $\mathbf B(\Gamma)$ are
concerned, we require that  
$| \left\langle \nabla \cdot {\mathbf B(\Gamma)} \right\rangle_E | < \infty$,
that
$0 < | \left\langle \nabla{\cal H}\cdot{\mathbf B(\Gamma)} \right\rangle_E | < \infty $,
and that 
$(\ln | \left\langle \nabla{\cal H}\cdot{\mathbf B(\Gamma)} \right\rangle|)/N \simeq 0$.
Given the dependence of $A(E)$ on the Hamiltonian, the family of vector fields
${\mathbf B(\Gamma)}$ that obey these criteria is not immediately obvious, 
and we have not attempted to develop a more general method of generating such vector 
fields. We simply note that the last condition allows all finite-order polynomials 
and bounded functions of the $p_i$ and $q_i$, as well as ratios of finite-order 
polynomials. However, it remains clearer from the canonical case (where the 
domain of integration does not depend on the Hamiltonian) whether such functions 
will obey the first two conditions.

\subsection{Microcanonical Periodic and MD Systems}
\label{sub:micro:period}

Let us now consider the necessary changes to the proof of
Eq.\ (\ref{EMGET:microcanonical}) in order for it to hold in a periodic
system. For periodic systems, the extent of $\Omega$ in the spatial
coordinates is no longer determined by boundary walls, but by the size
and shape of the primitive cell. If the primitive cell of the periodic
system is the same size and shape as the bounded system, then $\Omega$ 
will be the same in both cases. 

The difference between the bounded and periodic systems is that
particles cannot pass through the walls of the bounded system. This
implies that the energy at the walls is infinite, so that our surfaces
of constant energy lie entirely \emph{within} $\Omega$, and do not
pass through the boundary, which we denote $\partial\Omega$. This
assumption is implicit in our application of Gauss' Theorem.

In the periodic system, particles \emph{can} pass through the
``walls'' of our primitive cell, reappearing on the other side of the
cell. Therefore surfaces of constant energy \emph{can} (and do) pass
through $\partial\Omega$. Thus, when we use Gauss' Theorem, our
Gaussian surface consists not only of $A(E)$ and $A(E+h)$, but also
all points on $\partial\Omega$ whose energies lie between $E$ and
$E+h$. This extra term is of the form
\[
\lim_{h\rightarrow0} \frac{1}{h} \left[ 
\int\limits_{\partial\Omega(E+h)} 
    {\mathbf B(\Gamma)} \cdot \hat{\mathbf{n}}(\bbox{\Gamma}) \;{\sf
d}{A} - 
\int\limits_{\partial\Omega(E)} 
    {\mathbf B(\Gamma)} \cdot \hat{\mathbf{n}}(\bbox{\Gamma}) \;{\sf
d}{A} \right]
= \int\limits_{\partial A(E)} {\mathbf B(\Gamma)} \cdot
\hat{\mathbf{n}}(\bbox{\Gamma}) \;{\sf d}{L},
\]
where $\partial\Omega(E) = \partial\Omega \cap \Omega(E)$, ${\sf
d}{A}$ is the volume measure on $\partial\Omega$, $\partial A(E) =
\partial\Omega \cap A(E)$, and ${\sf d}{L}$ is the volume measure on
$\partial A(E)$. Note that $\hat{\mathbf{n}}(\bbox{\Gamma})$ does not
necessarily point in the same direction as $\nabla{\cal
H}(\bbox{\Gamma})$, since the walls of the primitive cell are not
determined by the energy surfaces. For
Eq.\ (\ref{EMGET:microcanonical}) to hold in every microcanonical
ensemble, we require, as a condition on ${\mathbf B(\Gamma)}$, that
\begin{equation}
\label{condition:pbc}
\int\limits_{\partial A(E)} {\mathbf B(\Gamma)} \cdot
\hat{\mathbf{n}}(\bbox{\Gamma}) \;{\sf d}{L} = 0 \quad\forall E.
\end{equation}
To determine which functions satisfy this criterion, we consider a
system where one of the particles is at one of the walls of the
primitive cell \cite{f1}, corresponding to a phase point
$\bbox{\Gamma}_a$. There is an equivalent system where this particle
is placed on the ``opposite'' wall of the primitive cell, represented
by $\bbox{\Gamma}_b$. It follows that $\hat{\mathbf
n}(\bbox{\Gamma}_a) = -\hat{\mathbf n}(\bbox{\Gamma}_b)$. Therefore,
since $\bbox{\Gamma}_a$ and $\bbox{\Gamma}_b$ must lie in the same
microcanonical ensemble, if ${\mathbf B(\Gamma}_a)={\mathbf
B(\Gamma}_b)$, then the criterion of Eq.\ (\ref{condition:pbc}) is
satisfied. Therefore any function which is periodic in the primitive
cell [ie such that, if $\bbox{\Gamma}_a$ and $\bbox{\Gamma}_b$
describe the same state, then ${\mathbf B(\Gamma}_a)={\mathbf
B(\Gamma}_b)$] will satisfy Eq.\ (\ref{EMGET:gen}) for periodic
systems. Note that this is a sufficient condition but not a necessary
one.

Finally, let us consider the MD microcanonical ensemble. This ensemble
represents the family of systems encountered during a constant energy
molecular dynamics simulation, where linear momentum is conserved. Let
$\Omega_{\text{MD}}$ be the set of allowed $\bbox{\Gamma}$ for such a
simulation. Clearly $\Omega_{\text{MD}}$ is smaller than $\Omega$,
which admits all possibilities for the total linear momentum. Phase
points $\bbox{\Gamma}$ with the same linear momentum in the
$x$-direction, say, all lie on the same (hyper)plane in $\Omega$, so
that $\Omega_{\text{MD}}$ is the intersection of $\Omega$ with the
three phase space planes which correspond to conservation of linear
momentum in each Cartesian direction. We denote their normal vectors
as $\mathbf{\hat{P}}_x$, $\mathbf{\hat{P}}_y$, and
$\mathbf{\hat{P}}_z$.

The entropy will still correspond to the phase space volume, except
that this volume is now $6N-4$ dimensional. However, we can only apply
Gauss' Theorem to the projection of the vector field ${\mathbf
B(\Gamma)}$ onto $\Omega_{\text{MD}}$. Alternatively we must select
${\mathbf B(\Gamma)}$ so that it lies entirely in
$\Omega_{\text{MD}}$. Such a ${\mathbf B(\Gamma)}$ must satisfy the
condition that ${\mathbf B(\Gamma)}\cdot\mathbf{\hat{P}}_\alpha = 0
\quad (\alpha=x,y,z)$. In this case, Eq.\ (\ref{EMGET:microcanonical})
will generate the correct temperature in the MD ensemble.

\subsection{Canonical Ensemble}
\label{sub:canon}

We now move on to a proof of Eq.\ (\ref{EMGET:gen}) in the canonical
ensemble, starting with the bounded case. We invoke Gauss' Theorem
over $\Omega(E)$ for an arbitrary vector field in phase space
${\mathbf B(\Gamma)} e^{-\beta{\cal H}(\bbox{\Gamma})}$ (where we
assume a finite, positive $\beta$), ie
\begin{eqnarray*}
\int\limits_{A(E)} e^{-\beta{\cal H}(\bbox{\Gamma})} {\mathbf
B(\Gamma)} \cdot \hat{\mathbf{n}}(\bbox{\Gamma}) \;{\sf d}{A_E} 
&=&
\int\limits_{\Omega(E)} \nabla\cdot \left( {\mathbf B(\Gamma)}
e^{-\beta{\cal H}(\bbox{\Gamma})} \right) \;{\sf d}{\Gamma} \\
&=& 
\int\limits_{\Omega(E)} e^{-\beta{\cal H}(\bbox{\Gamma})} \nabla\cdot
{\mathbf B(\Gamma)} \;{\sf d}{\Gamma} - 
\ \beta \!\! \int\limits_{\Omega(E)} e^{-\beta{\cal H}(\bbox{\Gamma})}
\nabla {\cal H}(\bbox{\Gamma}) \cdot {\mathbf B(\Gamma)} \;{\sf
d}{\Gamma}.
\end{eqnarray*}
In the limit as $E\rightarrow\infty$, we obtain
\begin{equation}
 \label{EMGET:can1}
\lim_{E\rightarrow\infty} \ e^{-\beta E} \! \int\limits_{A(E)}
{\mathbf B(\Gamma)} \cdot \hat{\mathbf{n}}(\bbox{\Gamma}) \;{\sf
d}{A_E}  = 
\int\limits_{\Omega} e^{-\beta{\cal H}(\bbox{\Gamma})} \nabla\cdot
{\mathbf B(\Gamma)} \;{\sf d}{\Gamma} - 
\beta \int\limits_{\Omega} e^{-\beta{\cal H}(\bbox{\Gamma})} \nabla
{\cal H}(\bbox{\Gamma}) \cdot {\mathbf B(\Gamma)} \;{\sf d}{\Gamma}.
\end{equation}
For Eq.\ (\ref{EMGET:can1}) to be of any use, we require that the two
integrals on the right hand side be finite. For the latter integral
this gives us
\[
\left| \int\limits_{\Omega} e^{-\beta{\cal H}(\bbox{\Gamma})} \nabla
{\cal H}(\bbox{\Gamma}) \cdot {\mathbf B(\Gamma)} \;{\sf d}{\Gamma}
\right| < \infty \quad\Rightarrow\quad
 \lim_{E\rightarrow\infty} \int\limits_{A(E)} e^{-\beta E} {\mathbf
B(\Gamma)} \cdot \hat{\mathbf{n}}(\bbox{\Gamma}) \;{\sf d}{A_E} = 0.
\]
This means that whenever the last integral in Eq.\ (\ref{EMGET:can1})
exists, the left hand side of Eq.\ (\ref{EMGET:can1}) must be
identically zero. It follows by rearrangement that
\begin{equation}
\label{EMGET:can2}
\frac{1}{kT} = \beta
=
\frac{ \int\limits_{\Omega} e^{-\beta{\cal H}(\bbox{\Gamma})}
\nabla\cdot {\mathbf B(\Gamma)} \;{\sf d}{\Gamma} }
     { \int\limits_{\Omega} e^{-\beta{\cal H}(\bbox{\Gamma})} \nabla
{\cal H}(\bbox{\Gamma}) \cdot {\mathbf B(\Gamma)} \;{\sf d}{\Gamma} }
=
\frac{ \left\langle \nabla\cdot {\mathbf B(\Gamma)} \right\rangle }
     { \left\langle \nabla {\cal H}(\bbox{\Gamma}) \cdot {\mathbf
B(\Gamma)} \right\rangle },
\end{equation}
in agreement with Eq.\ (\ref{EMGET:gen}). Since $\beta$ is finite, we 
have subsumed the first two conditions on ${\mathbf B(\Gamma)}$ into
the proof. The third is implicit in the convergence of 
$ \left\langle \nabla{\cal H}\cdot{\mathbf B(\Gamma)} \right\rangle_E $. 
For Eq.\ (\ref{EMGET:can2}) to hold, we require that the integral
\[
\ e^{-\beta E} \! \int\limits_{A(E)}
{\mathbf B(\Gamma)} \cdot \hat{\mathbf{n}}(\bbox{\Gamma}) \;{\sf
d}{A_E}
=
\ e^{\beta [TS(E)-E]} \left\langle {\cal
B}(\bbox{\Gamma}) \right\rangle_E 
=
\ e^{\beta [TS_{\cal B}(E)-E]}
\]
converge. When we consider 
that $e^{\beta [TS(E)-E]}$ also converges, but that $e^{S(E)}$ 
does not, we immediately obtain the third condition on 
$\mathbf B(\Gamma)$. Thus the conditions for Eq.\ (\ref{EMGET:gen}) 
to hold in the canonical ensemble are the same as those for the 
microcanonical ensemble.

\subsection{Canonical Periodic and MD Ensembles}
\label{sub:canon:period}

As with the microcanonical case, we must be careful in our application
of Gauss' Theorem to canonical systems with periodic boundary
conditions. In analogy with the microcanonical case, our Gaussian
surface consists not only of $A(E)$, but of $\partial\Omega(E)$ as
well, and the left hand side of Eq.\ (\ref{EMGET:can1}) becomes
\[
\lim_{E\rightarrow\infty} \int\limits_{A(E)\cup\partial\Omega(E)}
e^{-\beta{\cal H}(\bbox{\Gamma})} {\mathbf B(\Gamma)} \cdot
\hat{\mathbf{n}}(\bbox{\Gamma}) \;{\sf d}{A_E}.
\]
We have already seen that the integral over $A(E)$ must go to zero in
order for $\left\langle \nabla {\cal H}(\bbox{\Gamma}) \cdot {\mathbf
B(\Gamma)} \right\rangle$ to exist, so we simply require that
\[
\int\limits_{\partial\Omega} e^{-\beta{\cal H}(\bbox{\Gamma})}
{\mathbf B(\Gamma)} \cdot \hat{\mathbf{n}}(\bbox{\Gamma}) \;{\sf d}{A}
= 0 \quad\forall \beta.
\]
However, via the properties of the Laplace transform we have that
\[
\int\limits_{\partial\Omega} e^{-\beta{\cal H}(\bbox{\Gamma})}
{\mathbf B(\Gamma)} \cdot \hat{\mathbf{n}}(\bbox{\Gamma}) \;{\sf d}{A}
= 0 \quad\forall \beta \quad \Leftrightarrow 
\int\limits_{\partial A(E)} {\mathbf B(\Gamma)} \cdot
\hat{\mathbf{n}}(\bbox{\Gamma}) \;{\sf d}{L} = 0 \quad\forall E.
\]
Therefore the condition under which Eq.\ (\ref{EMGET:gen}) will hold in
all canonical ensembles is equivalent to the condition under which
Eq.\ (\ref{EMGET:gen}) will hold in all microcanonical ensembles.
Just as in the microcanonical case, Eq.\ (\ref{EMGET:can2})
will hold in the canonical ensemble as long as ${\mathbf B(\Gamma)}$
is periodic in $\Omega$.

Finally, we consider the canonical MD ensemble. As with the
microcanonical MD ensemble, our application of Gauss' Theorem requires
that ${\mathbf B(\Gamma)}$ lie in $\Omega_{\text{MD}}$, so we again
require that ${\mathbf B(\Gamma)}\cdot\mathbf{\hat{P}}_\alpha = 0
\quad (\alpha=x,y,z)$ for Eq.\ (\ref{EMGET:can2}) to hold in the
canonical MD ensemble.

Thus the conditions for Eq.\ (\ref{EMGET:gen}) to hold for the 
periodic boundary system and the ``MD ensembles'' are the same 
in both the canonical and microcanonical ensembles.

\section{Formulae}
\label{sec:formulae}

Having proven Eq.\ (\ref{EMGET:gen}) in the canonical and
microcanonical ensembles, and found the conditions for it to hold in
systems with periodic boundary conditions and the MD ensembles, we now
demonstrate its use in generating expressions whose phase space
average yields the system temperature.

If we choose ${\mathbf B(\Gamma)}=(0,\ldots,\Gamma_i,\ldots,0)$, so
that only the $i$-th component is non-zero, then we obtain
\begin{equation}
\label{equip}
kT_{\text{equip}} = \frac{ \left\langle \nabla {\cal H}(\bbox{\Gamma})
\cdot {\mathbf B(\Gamma)} \right\rangle }
     { \left\langle \nabla\cdot {\mathbf B(\Gamma)} \right\rangle }  =
\left\langle \Gamma_i \frac{\partial {\cal H}}{\partial \Gamma_i}
\right\rangle
\end{equation}
This is the familiar Generalised Equipartition Theorem. If $\Gamma_i$
is a momentum, then we obtain the Equipartition Theorem, $\left\langle
p_i^2/m \right\rangle =kT$. If it is a coordinate, then we obtain the
lesser known Clausius Virial Theorem, $\left\langle -q_i F_i
\right\rangle =kT$, where $F_i$ is the generalised force acting on
coordinate $q_i$ \cite{munster,tolman}. We note that the Clausius
Virial Theorem gives a function of coordinates only, whose average is
the temperature of the system. However, the function ${\mathbf
B(\Gamma)}\left(=\bbox{\Gamma}\right)$ is not periodic in $\Omega$ in
this case, so that this theorem does \emph{not} hold for periodic
systems. It is therefore of little use to practitioners of most MD
simulations as a means of calculating the temperature.

If we select an arbitrary vector field ${\mathbf X(\Gamma)}$, and
choose
\begin{equation}
\label{rugh:vfield}
{\mathbf B(\Gamma)} = \frac{{\mathbf X(\Gamma)}}{\nabla{\cal
H}(\bbox{\Gamma})\cdot {\mathbf X(\Gamma)}},
\end{equation}
then for \emph{all} choices of ${\mathbf X(\Gamma)}$, ${\cal
B}(\bbox{\Gamma}) \equiv 1$. Consequently, we obtain
\begin{equation}
\label{rugh:gen}
\frac{1}{kT} = \left\langle \nabla\cdot\frac{{\mathbf
X(\Gamma)}}{\nabla{\cal H}(\bbox{\Gamma})\cdot {\mathbf X(\Gamma)}}
\right\rangle,
\end{equation}
providing this average exists.
 Substituting ${\mathbf X(\Gamma)}=\nabla{\cal H}(\bbox{\Gamma})$, we
obtain Rugh's final equation \cite{rugh}. Since the Hamiltonian is
periodic in systems with periodic boundary conditions, ${\mathbf
B(\Gamma)}$ will also be periodic, so that Rugh's result holds in
periodic systems. Furthermore, it satisfies the criterion for the MD
ensembles, so that it can be applied to MD simulations as well.

\section{Example : Simulation Application}
\label{sec:example}

In this section we consider the application of Eq.\ (\ref{EMGET:gen})
to a simulation of a system of particles interacting with a
short-range pair potential, as in \cite{brent,gary}. These simulations
employ periodic boundary conditions, and as a consequence, the forces
acting on a body are not correlated with the \emph{absolute} positions
of the particles, but only their relative positions. Thus many of the
simple vector fields whose divergences are easily calculated (such as
${\mathbf B(\Gamma)} = {\mathbf \Gamma }$) do not satisfy the first 
criterion for Eq.\ (\ref{EMGET:gen}) to hold; in this case,
$\left\langle \nabla{\cal H}\cdot{\mathbf \Gamma} \right\rangle_E = 0$.
In general, it is more difficult to find functions which are correlated
with the inter-particle forces. Due to this difficulty, from this point
on we restrict ourselves to choices of ${\mathbf B(\Gamma)}$ which are 
directly related to $\nabla{\cal H}$, to ensure that this condition is 
met.

\subsection{Theory}

In  systems of particles interacting with a short-range pair potential
$\Phi(r)$, the Hamiltonian can be separated into a momentum
contribution (the kinetic energy $K$) and a spatial contribution (the
potential energy $V$), ie
\begin{equation}
\label{ha:gen}
{\cal H} ({\mathbf \Gamma}) = K(\{p_i\}) + V(\{q_i\})= \sum_{i=1}^{3N}
\frac{p_i^2}{2m} + 
\sum_{i=1}^N \sum_{j<i} \Phi(\|\bbox{r}_{ij}\|),
\end{equation}
where $\bbox{r}_{ij} = \bbox{r}_i - \bbox{r}_j$, and $\bbox{r}_i$ is
the vector describing the position of the $i$-th particle. If the
potential has a continuous first derivative, then $\nabla{\cal H}$
satisfies the requirements of Gauss' Theorem, and consequently those
of our temperature expressions. Note that if we define $\bbox{r}_{ij}$
as the \emph{minimum image} separation of the $i$-th and $j$-th
particles, then $V$ is periodic in the spatial coordinates. Thus
$\nabla{\cal H}$ will be periodic as well. As a consequence, we can
obtain the temperature of our system using Rugh's expression, ie by
substituting ${\mathbf X(\Gamma)}=\nabla{\cal H}(\bbox{\Gamma})$ into
Eq.\ (\ref{rugh:gen}) above.

We now make the following important observation --- since $\nabla{\cal
H}$ satisfies the criteria for Eq.\ (\ref{EMGET:gen}) to hold in
periodic boundary systems and MD ensembles, it follows that $\nabla K$
and $\nabla V$ must as well. Therefore, we would expect to be able to
generate the temperature by substituting 
${\mathbf X(\Gamma)}=\nabla K(\{p_i\})$ 
and 
${\mathbf X(\Gamma)}=\nabla V(\{q_i\})$ 
into Eq.\ (\ref{rugh:gen}).

Since the interaction potential is short-ranged, $V$ will grow as $N$
in the thermodynamic limit. Consequently, the Hamiltonian grows as $N$
in the thermodynamic limit, and if we substitute 
${\mathbf B(\Gamma)}=\nabla{\cal H}(\bbox{\Gamma})$,
${\mathbf B(\Gamma)}=\nabla K(\{p_i\})$ and 
${\mathbf B(\Gamma)}=\nabla V(\{q_i\})$ into Eq.\ (\ref{EMGET:gen}),
we would also expect to generate the temperature.

In this paper we will not examine the temperatures generated from the
kinetic energy, since they are closely related to the equipartition
temperature [Eq.\ (\ref{equip})], and do not reveal any new results.
Our interest lies in the fact that temperature expressions generated
with $\nabla V(\{q_i\})$ contain \emph{no} explicit reference to the
momenta in our system, a fact which has been exploited in
\cite{brent}. The temperature we obtain from substituting ${\mathbf
X(\Gamma)}=\nabla{\cal H}(\bbox{\Gamma})$ into Eq.\ (\ref{rugh:gen}) we
denote by $T_{\text{norR}}$ --- ``nor'' since it is generated using
the normal vector field $\nabla{\cal H}$, and ``R'' since it is
generated using Rugh's prescription. In a similar manner, we denote by
$T_{\text{conR}}$ the temperature we obtain from substituting
${\mathbf X(\Gamma)}=\nabla V(\{q_i\})$ --- the configurational part
of the Hamiltonian --- into Eq.\ (\ref{rugh:gen}). When substituting
these vector fields into Eq.\ (\ref{EMGET:gen}), we denote the
corresponding temperatures as $T_{\text{norF}}$ and $T_{\text{conF}}$,
``F'' denoting that we are calculating a ratio (fraction) of averages
in this case.

In making the appropriate substitutions, we obtain the following
expressions :
\begin{mathletters}
\label{ts}
\begin{equation}
\frac{1}{kT_{\text{norR}}} = 
\left\langle 
   \frac{   \frac{3N}{m} - \sum_i \nabla_i \cdot {\mathbf F}_i }
        {   \sum_i  \frac{{\mathbf p}_i^2}{m^2} + {\mathbf F}_i ^2  }
  -
        \frac{  2 \sum_i\frac{{\mathbf p}_i^2}{m^3} + 2 \sum_{ij}
{\mathbf F}_i {\mathbf F}_j : \nabla_{i}{\mathbf F}_j  }
             {  (\sum_i  \frac{{\mathbf p}_i^2}{m^2} + 
{\mathbf F}_i ^2 )^2}  \right\rangle
,   \label{tnorr}
\end{equation}
\begin{equation}
\frac{1}{kT_{\text{conR}}} = 
\left\langle 
       \frac{ - \sum_i \nabla_i \cdot {\mathbf F}_i  }
            {   \sum_i  {\mathbf F}_i ^2       }  -
       \frac{ 2 \sum_{ij} {\mathbf F}_i {\mathbf F}_j :
\nabla_{i}{\mathbf F}_j  }
            {  (\sum_i  {\mathbf F}_i ^2)^2       }  \right\rangle
,   \label{tconr}
\end{equation}
\begin{equation}
\frac{1}{kT_{\text{norF}}} = 
\frac{  \left\langle   \frac{3N}{m} - \sum_i \nabla_i \cdot {\mathbf
F}_i  \right\rangle     }
     {  \left\langle   \sum_i \frac{{\mathbf p}_i^2}{m^2} + {\mathbf
F}_i^2   \right\rangle   }
,     \label{tnorf} 
\end{equation} 
\begin{equation}
\frac{1}{kT_{\text{conF}}} = 
\frac{  \left\langle   - \sum_i \nabla_i \cdot {\mathbf F}_i
\right\rangle }
     {  \left\langle     \sum_i  {\mathbf F}_i ^2           
\right\rangle }
,      \label{tconf} 
\end{equation}
\end{mathletters}
where the label $i$ refers to the \emph{particle}, rather than the
generalised coordinate. ${\mathbf F}_i$ represents the (vector) force
acting on particle $i$, ${\mathbf p}_i$ represents its momentum,
$\nabla_i = [\frac{\partial }{\partial x_i},\frac{\partial }{\partial
y_i},\frac{\partial }{\partial z_i}]$, where $x_i,y_i$ and $z_i$ refer
to the Cartesian coordinates of ${\mathbf r}_i$, and $:$ represents
the dyadic operator (ie for vectors $\mathbf{a},\mathbf{b}$ and matrix
$\mathbf{M}$, $\mathbf{a} \mathbf{b} : \mathbf{M} =
\sum_{\alpha,\beta} \mathbf{a}_{\alpha} \mathbf{b}_{\beta}
\mathbf{M}_{\beta\alpha}$). Eq.\ (\ref{tconf}) corresponds to the
temperature expression used in \cite{brent}, and Eq.\ (\ref{tnorf})
corresponds to the temperature expression used in \cite{gary}.

If we consider the second term on the right hand side of
Eqs.\ (\ref{tnorr},\ref{tconr}), the numerator increases as $N$ for a
short-ranged potential (since ${\mathbf F}_i {\mathbf F}_j :
\nabla_{i}{\mathbf F}_j$ will not contribute anything at large
particle separations), but the denominator increases as $N^2$. Therefore
this second term becomes negligible in the thermodynamic limit. Thus
the order 1 term is contained in the first term on the right hand
side of Eqs.\ (\ref{tnorr},\ref{tconr}). We will denote by 
$T_{\text{nor1}}$ and $T_{\text{con1}}$ the temperature calculated by 
the omission of these second terms respectively, ie
\begin{mathletters}
\label{ts2}
\begin{equation}
\frac{1}{kT_{\text{nor1}}} = 
\left\langle 
        \frac{   \frac{3N}{m} - \sum_i \nabla_i \cdot {\mathbf F}_i   
}
             {   \sum_i  {\mathbf F}_i ^2 + \frac{{\mathbf
p}_i^2}{m^2}   }  \right\rangle ,
\eqnum{14e}\label{tnor1} 
\end{equation}
\begin{equation}
\frac{1}{kT_{\text{con1}}} = 
\left\langle 
       \frac{ - \sum_i \nabla_i \cdot {\mathbf F}_i  }
            {   \sum_i  {\mathbf F}_i ^2       }          
\right\rangle .
\eqnum{14f}\label{tcon1}  \addtocounter{equation}{-1}
\end{equation}
\end{mathletters}

We expect that the temperatures given by
Eqs.\ (\ref{tnorr})--(\ref{tcon1}) should all be equal in the
thermodynamic limit. It is therefore of interest to compare their
rates of convergence to this limit, in order to ascertain the
appropriateness of their use.

\subsection{Results}

As an application of the above theory, we considered a
three-dimensional microcanonical WCA-potential system. The WCA
potential is defined as follows \cite{bible}, 
\begin{equation}
\label{defn:WCA}
\Phi(r)= \left\{ \begin{array}{ll}
4 \epsilon \left[ \left( \frac{\sigma}{r} \right)^{12} - \left(
\frac{\sigma}{r} \right)^6  \right] + \epsilon , &r < 2^{1/6} \sigma
\\
0, & \text{otherwise}
\end{array} \right. ,
\end{equation}
where $\sigma$ and $\epsilon$ represent our units of length and energy
respectively. This potential is continuous, has a continuous first
derivative and a piecewise continuous second derivative. Due to the 
discontinuity in the derivative of the force, errors appear in the 
computed system trajectories whenever the separation between two particles
crosses the $r=2^{1/6}\sigma$ boundary. However, these errors are too
small, in comparison with system size errors, to affect our results. 
Thus, if we substitute this pair potential into Eq.\ (\ref{ha:gen}), 
then we expect each of the temperatures defined in 
Eqs.\ (\ref{tnorr})--(\ref{tcon1}) to be equal, to order $(\ln N)/N$.

Values of these six temperatures were calculated for the 3D
microcanonical simulation of a periodic WCA system at various sizes
$N$, reduced number densities $\rho^*$, and reduced total energies 
per particle $\bar{E}^*$. They were determined by the average
of ten separate simulations, each of 200000 timesteps (of $\delta
t^*=0.001$). The errors associated with each temperature were given by
one third of the maximum deviation from the average over these ten
runs.

The first comparison was made between systems with the same density
and size, but differing energies. The values of the equipartition and
normal temperatures ($T_{\text{equip}}$, $T_{\text{norR}}$,
$T_{\text{norF}}$ and $T_{\text{nor1}}$) were calculated for a system
of 500 particles with reduced density $\rho^* = 0.8$, and reduced
energies per particle ranging from $\bar{E}^*=0.8$ to $\bar{E}^*=2.5$.
These values appear in Table \ref{energies}. The four temperatures
agree to within 0.6--0.8\% of the equipartition temperature over the
range of energies shown.

The values for the three configurational temperatures match the
corresponding normal temperatures to within 0.01\%, ie to the number
of digits shown in Table \ref{energies}. This can be explained in
terms of the kinetic and configuration terms in the numerator and
denominator of the normal temperature expressions. At high densities,
the configuration terms are much larger than those contributed by the
momentum terms --- in two dimensions, this is typically a difference
of four orders of magnitude, and in three dimensions, the difference
is about six orders of magnitude. For this reason, the value of the
normal temperature can be considered as a ``perturbation'' to the
corresponding configurational temperature which has a negligible
effect on our results. It is interesting to note, given this
dependence on the physical structure of the system rather than on its
momentum distribution, that the normal and configurational temperature
expressions yield the correct temperature across the solid-liquid phase
transition, despite the difference in the microscopic arrangements of 
atoms on either side of the transition temperature.

\begin{figure}
\psfig{figure=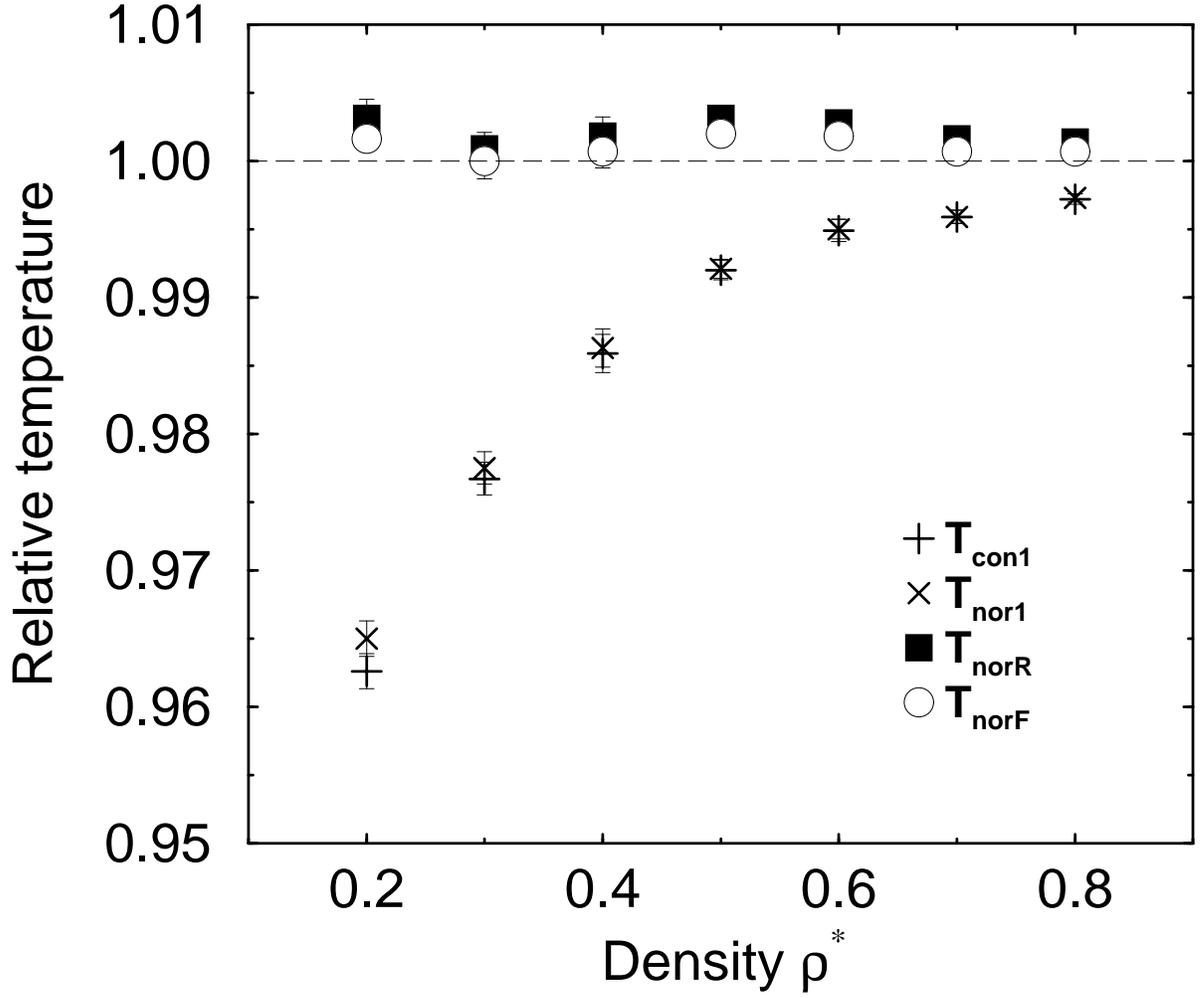,angle=270,width=\textwidth}
\vspace{2cm}
\caption{Variation of temperature values with system density. For a
system of 864 particles at reduced energy per particle
$\bar{E}^*=1.5$, various temperatures are given for different reduced
densities. Temperatures are reported as a fraction of the
equipartition temperature. }
\label{densities}
\end{figure}

In Fig.\ \ref{densities}, we compare a series of systems of fixed
energy per particle ($\bar{E}^*=1.5$) and system size ($N=864$), but with 
varying densities. The discrepancy between $T_{\text{nor1}}$ and
$T_{\text{norR}}$ increases when the density of the system is
decreased --- while $T_{\text{norR}}$ and $T_{\text{norF}}$ agree
quite well with the equipartition values, $T_{\text{nor1}}$ becomes
less and less reliable. However, in the thermodynamic limit,
$T_{\text{nor1}}$ must converge to the other two normal temperatures.
This result indicates that, while $T_{\text{nor1}}$ and $T_{\text{norR}}$ 
must converge towards the thermodynamic temperature, irrespective of 
the density, larger systems sizes are required for the same degree of 
convergence of $T_{\text{nor1}}$ as the density drops.

We should also note from Fig.\ \ref{densities} that, while
$T_{\text{norR}}$ and $T_{\text{norF}}$ are indistinguishable from
their configurational counterparts on the scale of the graph (and
hence are not shown), the difference between $T_{\text{nor1}}$ and
$T_{\text{con1}}$ becomes evident below densities of 
$\rho^*\approx0.5$. This is a result of the drop in the number of
particle interactions per timestep at lower densities. When the number
of these interactions is reduced, the configurational contributions do
not dominate the kinetic contributions as they do in the high density
regime. Consequently, the inclusion of kinetic terms (which, by
themselves would produce a value within 0.1\% of the equipartition
value) in $T_{\text{nor1}}$ will always correct $T_{\text{con1}}$
towards the equipartition value.

\begin{figure}
\psfig{figure=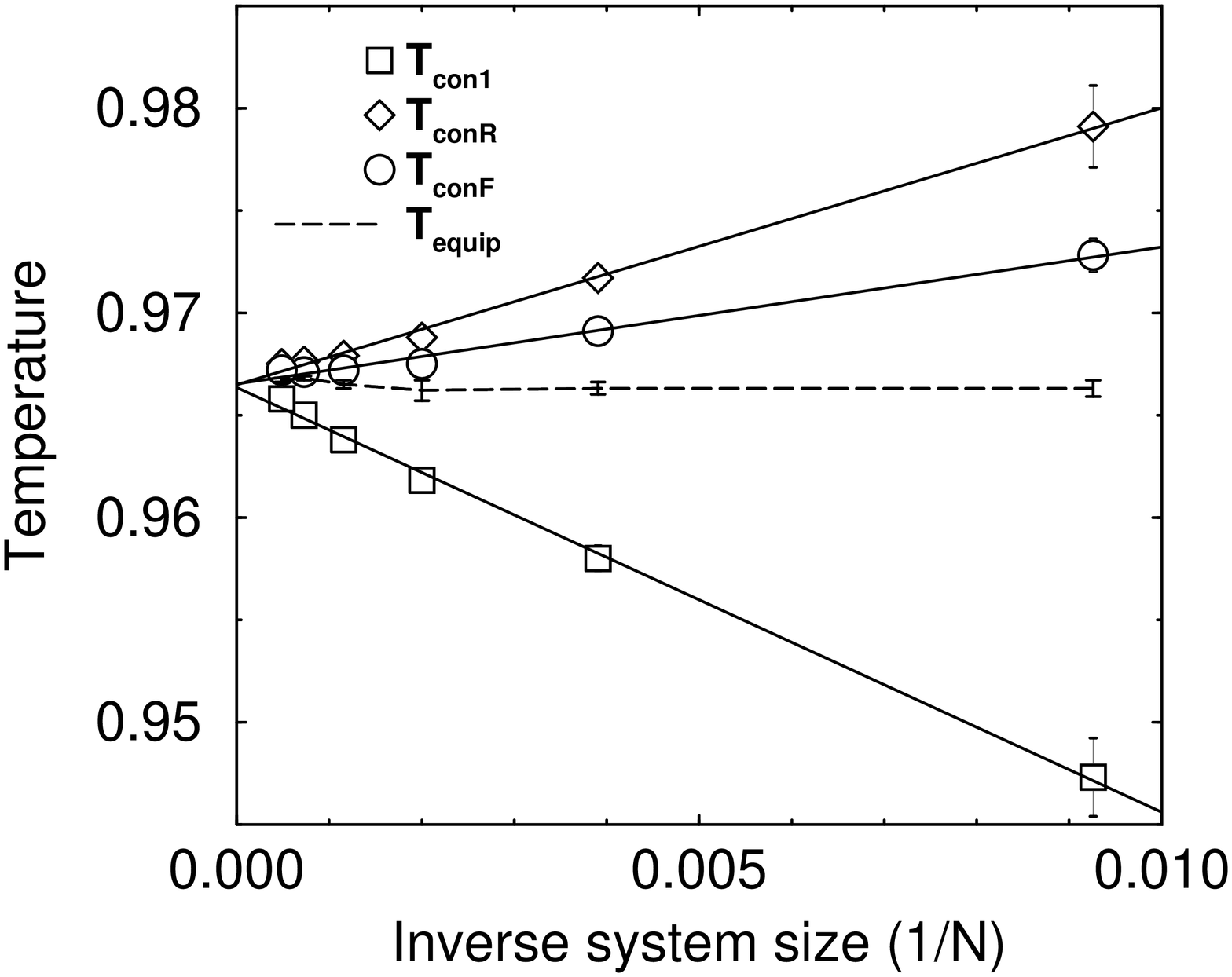,angle=270,width=\textwidth}
\vspace{2cm}
\caption{Variation of temperature values with system size. For the
state point $\rho^*=0.8$, $\bar{E}^*=1.5$, the three configurational
temperatures and the equipartition temperature are given for different
system sizes.}
\label{sizes}
\end{figure}

To further examine the system size dependence of our temperature
expressions, we consider a single state point ($\rho^*=0.8,
\bar{E}^*=1.5$), and compare the temperature expressions as a function
of the number of particles in the system, ranging from $N=108$ to
$N=2048$. The results of this comparison appear in Fig.\ \ref{sizes},
where the three configurational temperatures have been plotted against
inverse system size. At this density, the difference between the
normal temperatures and the corresponding configurational temperatures
is not distinguishable on the scale of the graph for all but the 108
particle system (where the discrepancy is 0.02\%), so we show only the
configurational temperatures. We observe, within the errors of our
calculations, the convergence of all four temperatures 
towards a common value. We would interpret this value as the
thermodynamic temperature of a system at that state point, in the
thermodynamic limit.

\section{Conclusion}
\label{sec:conc}

We have derived a general functional which, given a vector field 
${\mathbf B(\Gamma)}$ which satisfies certain broad conditions, will 
determine the thermodynamic temperature of an equilibrium system 
in the thermodynamic limit via Eq.\ (\ref{EMGET:gen}). Its rate of 
convergence in the thermodynamic limit will be determined by the order 
of $\left\langle \mathbf B(\Gamma)\cdot\nabla{\cal H} \right\rangle$. We 
note, however, that if we define ${\mathbf B(\Gamma)}$ as per
Eq.\ (\ref{rugh:vfield}), then $\left\langle {\mathbf B(\Gamma)}\cdot\nabla{\cal
H} \right\rangle \equiv1$, and what we obtain in Eq.\ (\ref{EMGET:gen})
is precisely the derivative of the logarithm of the ensemble phase space 
volume with respect to the energy. In the thermodynamic limit, this will 
yield the thermodynamic temperature $\partial S/\partial E$. However, 
for different ${\mathbf B(\Gamma)}$, the value we obtain will depend upon our 
sampling of phase space during the simulation, and hence the values 
obtained from different expressions may vary. The temperature expressions 
$T_{\text{norR}}$ and $T_{\text{conR}}$ fall into this category.

One practical problem that arises from the application of
Eq.\ (\ref{EMGET:gen}) to periodic boundary systems is the
difficulty in avoiding vector fields ${\mathbf B(\Gamma)}$ such that
 $\left\langle {\mathbf B(\Gamma)}\cdot\nabla{\cal H}\right\rangle=0$. 
To circumvent this problem, we have only considered vector fields
${\mathbf B(\Gamma)}$ that are linear transformations of $\nabla{\cal H}$.
This approach is by no means exhaustive, but serves to demonstrate
one application of this theory.

It is clear from Eqs.\ (\ref{tnorr})--(\ref{tcon1}) that
$T_{\text{norR}}$ will be computationally more expensive than
$T_{\text{nor1}}$ or $T_{\text{norF}}$ --- the omitted term involves
calculations which assume the intermolecular forces to have already
been evaluated, thus requiring a second force loop. It is therefore of
interest to determine whether these approximations to $\partial S/\partial E$ 
make a useful substitution for $T_{\text{norR}}$. From our results we 
conclude that $T_{\text{norF}}$ is more reliable than $T_{\text{nor1}}$, and in
our work is a useful expression for the temperature whenever $T_{\text{norR}}$ 
is valid. It is for these reasons that the fractional forms ($T_{\text{norF}}$ or
$T_{\text{conF}}$) appear in \cite{brent,gary}.

Eq.\ (\ref{EMGET:gen}) has important consequences for practitioners of
non-equilibrium MD simulations, stressing the fact that the
instantaneous kinetic energy per kinetic degree of freedom is
\emph{not} the only function whose ensemble average yields the
temperature. The preference given to the kinetic energy is generally
due to its ease of calculation: apart from this, there is no reason
--- in both equilibrium and non-equilibrium calculations --- to prefer
the kinetic energy expression over any other.

\begin{table}
\caption{A comparison of values of the three normal temperatures with
values of the equipartition temperature, for simulations of systems
with size $N=500$, reduced density $\rho^*=0.8$, and various reduced
energies per particle $\bar{E}^*$. For each normal temperature, two
values are reported. The first is the temperature as determined from
the simulation, and the second is the discrepancy between that normal
temperature and the equipartition temperature, given as a percentage
of the equipartition temperature. The numbers in brackets indicate the
error in the last decimal place given.}
\begin{tabular}{cccccccc}
$\bar{E}^* $ & $T_{\text{equip}}$ & \multicolumn{2}{c}{$T_{\text{norR}}$} &
\multicolumn{2}{c}{$T_{\text{norF}}$} &
\multicolumn{2}{c}{$T_{\text{nor1}}$}  \\
&& abs & rel (\%) & abs & rel (\%) & abs & rel (\%) \\
\hline
 0.8 & 0.5081 (1) & 0.5097 (3) & 0.31 & 0.5090 (1) & 0.19 & 0.5065 (3)
& -0.31 \\
 1.0 & 0.6374 (1) & 0.6389 (7) & 0.24 & 0.6381 (4) & 0.11 & 0.6348 (7)
& -0.41 \\
 1.2 & 0.7679 (2) & 0.7705 (4) & 0.34 & 0.7694 (1) & 0.20 & 0.7652 (4)
& -0.35 \\
 1.5 & 0.9664 (2) & 0.9701 (6) & 0.38 & 0.9687 (2) & 0.24 & 0.9631 (6)
& -0.34 \\
 1.8 & 1.1671 (3) & 1.1709 (6) & 0.32 & 1.1694 (3) & 0.20 & 1.1621 (6)
& -0.43 \\
 2.0 & 1.3024 (2) & 1.3060 (6) & 0.28 & 1.3042 (3) & 0.14 & 1.2958 (6)
& -0.51 \\
 2.2 & 1.4386 (2) & 1.4424 (9) & 0.26 & 1.4403 (9) & 0.12 & 1.4307 (9)
& -0.55 \\
 2.5 & 1.6450 (2) & 1.6497 (9) & 0.28 & 1.6473 (8) & 0.14 & 1.6359 (9)
& -0.55 \\
\end{tabular}
\label{energies}
\end{table}
\newpage

\newpage

\begin{references}
\bibitem[*]{f1}
If two or more particles are at the wall, or if the particle is in a
corner, then this corresponds to a subset of $\partial\Omega$ of
measure zero.
\end{references}
\end{document}